%% file: main.tex
\pdfoutput=1

\documentclass[11pt]{article}
\newcommand{\comment}[1]{} 
\usepackage[]{EACL2023}

\usepackage{times}
\usepackage{latexsym}

\usepackage[T1]{fontenc}

\usepackage[utf8]{inputenc}

\usepackage{microtype}
\usepackage{bbm}
\usepackage{xcolor}
\usepackage[toc,page]{appendix}
\usepackage{hyperref}
\usepackage{url}
\usepackage{graphicx}
\usepackage{booktabs}
\usepackage{graphics}
\usepackage{tabularx}
\usepackage{xspace}
\usepackage{graphicx}
\usepackage{booktabs}
\usepackage{mdwlist}
\usepackage{multirow}
\usepackage{amsmath}
\usepackage{mathtools}
\usepackage{enumitem}
\usepackage{calc}
\usepackage[neverdecrease]{paralist}
\usepackage[ruled,noend]{algorithm2e}
\usepackage{booktabs}
\usepackage{diagbox}
\usepackage{colortbl}
\usepackage{subcaption}
\usepackage{adjustbox}
\usepackage{makecell}
\usepackage{verbatim}
\usepackage{amsmath}
\usepackage{color,soul}
\usepackage[ruled,noend]{algorithm2e}
\usepackage{booktabs}
\usepackage{diagbox}
\usepackage{colortbl}
\usepackage{subcaption}
\usepackage{makecell}
\usepackage{verbatim}
\usepackage{amsmath}
\usepackage{silence}
\usepackage{stfloats}
\usepackage[english]{babel}

\title{Intent Identification and Entity Extraction for Healthcare Queries in Indic Languages}

 \author{$^1$Ankan Mullick \qquad $^2$Ishani Mondal\thanks{\hspace{1mm} Authors contributed equally} \qquad $^1$Sourjyadip Ray$^*$\qquad $^3$R Raghav \qquad \\{\bf $^1$G Sai Chaitanya \and $^1$Pawan Goyal} \\ \texttt{\{$^1$ankanm, $^1$sourjyadipray\}@kgpian.iitkgp.ac.in, $^2$imondal@umd.edu} \\ \texttt{$^3$rraghavr@cs.cmu.edu, \{$^1$gajulasai@, $^1$pawang@cse.\}iitkgp.ac.in}\\ $^1$Computer Science and Engineering Department, IIT Kharagpur, India \\ $^2$University of Maryland, College Park, USA \\ $^3$School of Computer Science, Carnegie Mellon University, USA}

\begin{document}
\maketitle
\begin{abstract}
Scarcity of data and technological limitations for resource-poor languages in developing countries like India poses a threat to the development of sophisticated NLU systems for healthcare.
To assess the current status of various state-of-the-art language models in healthcare, this paper studies the problem by initially proposing two different Healthcare datasets, Indian Healthcare Query Intent-WebMD and 1mg (IHQID-WebMD and IHQID-1mg) and one real world Indian hospital query data in English and multiple Indic languages (Hindi, Bengali, Tamil, Telugu, Marathi and Gujarati) which are annotated with the query intents as well as entities. 
Our aim is to detect query intents and extract corresponding entities. We perform extensive experiments on a set of models in various realistic settings and explore two scenarios based on the access to English data only (less costly) and access to target language data (more expensive). We analyze context specific practical relevancy through empirical analysis. 
The results, expressed in terms of overall F1 score show that our approach is practically useful to identify intents and entities. 
\end{abstract}

\section{Introduction}

Healthcare is a top priority for every country. 
People across the world ask millions of health-related queries, hoping to get a response from a domain expert \cite{doi:10.1200/GO.20.00118}. 
These queries mostly deal with medical history of patients, possible drug interactions, disease related concerns, treatment protocols and so on. 
Conversational agents for healthcare play a pivotal role by facilitating useful information dissemination \cite{li-etal-2020-jennifer, maniou2020employing}.
In order to understand these queries better, practical conversational systems for healthcare need to be developed.
However, the primary obstacle in developing such technologies for low-resource languages is the lack of usable data \cite{mehta-etal-2020-learnings,daniel-etal-2019-towards,effectivetransfer}.

India is a country with a diverse language speaking population suffering from abject poverty and low-economic status \cite{Mohanty+2010+131+154,PANDE20032075}.
This linguistic diversity and complex socio-economic situation in India
certainly poses significant challenges in developing automatic healthcare systems; and there is a lack of linguistic resources specific to the medical domain. For example, situations such as the patient and the doctor speaking in different languages, is not an uncommon situation in rural India. These individuals are unable to avail the existing systems and facilities which exist mainly in the English language. 
Recent efforts in developing automatic translation systems, even from extremely low resource languages such as `Mundari' and `Gondi' \cite{joshi2019unsung}, should ideally improve this situation, but there is no extensive study on that front. 

\begin{figure*}[!htp]
    \centering
    \begin{adjustbox}{width=0.90\linewidth}
    \includegraphics[width=\linewidth]{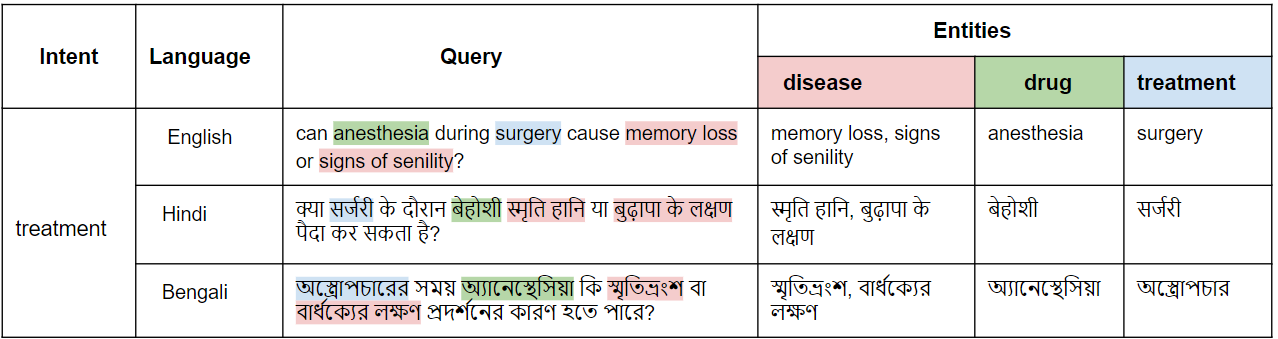}
    \end{adjustbox}
    \caption{Example of a query of `treatment' intent category for different languages along with associated entities.}
    \label{fig:example-intro}
\end{figure*}

In order to bridge this language barrier, massively Multilingual Transformer based Language Models (MMLM) \cite{devlin2019bert, lample2019cross} have made impressive advancements on a wide range of downstream applications.
But the real-world implications of such advancements in the Indian healthcare system remain largely unexplored.
In this paper, we aim to explore scarcity of the data and study the extent to which the existing language technologies can be leveraged to develop practically useful healthcare systems for the low-resource languages in developing countries.

With an aim to answer our research question, we create two different multilingual healthcare datasets, namely, IHQID-WebMD and IHQID-1mg. These datasets are created by crawling frequently asked questions from two healthcare websites, \textit{WebMD} and \textit{1mg}. 
These datasets comprise frequently asked questions about drugs, diseases and treatment methods in seven different languages, namely English, Hindi, Bengali, Tamil, Telugu, Gujarati and Marathi.
The queries are manually tagged with intent labels and entity tags by domain-experts and translated by native speakers of the corresponding languages.
We also collect real world Indian hospital queries (annotated) in seven languages to check the 
empirical effectiveness of our approach. 
Fig. \ref{fig:example-intro} shows an example of a health query belonging to `treatment' intent class manually translated into three different languages. 
Then we evaluate the performance of state-of-the-art language models (LMs), for both English and multilingual setups on our datasets, to answer the questions regarding their deployability and practicality. Various experimental configurations (Section 4) have been tried on these datasets where we try to figure out the ways of using best technologies through extensive experimentation in two real-world scenarios. First, we assume to have access to only English training queries (less costly) and the test queries are multilingual in nature. We observe that translate-test setup on RoBERTa seems to be a reasonable choice of technology. Second, we assume to have access to manually written multilingual training and test queries in the target languages, which is indeed quite expensive in terms of data collection effort. However, back-translation of both train and test queries proves to be a reasonable choice if we have budget of collecting data in target languages.

In sum, our contributions are 
four folds:\\
    
    - We propose two intent and entity labelled Indian healthcare datasets (annotated by domain-experts) comprising of frequently asked questions from users.
\\

    - Even though the large language models have proved their effectiveness in almost every NLU operation, we want to determine their effectiveness in determining the correct intent and slot filling operations for practical domain-specific healthcare scenarios in the Indian context. We intend to analyze how should we prioritize the research and resource building investments for the economically backward countries with a high percentage of multilingual population? This will make us aware about the best techniques of deploying the language models in various scenarios such as: availability of English training data vs multilingual training data. Keeping this in mind, all our experiments have been carried out using both monolingual and multilingual setups of these models. Through our experiments, we try to point out the best possible language models and techniques to develop practically useful NLU solutions (pipeline based approach for intent detection and corresponding entity extraction from the queries).
    \\
    
    - Through extensive experiments on the datasets, we recommend the community to use back-translation of test queries to English in two real-life scenarios as a reasonable choice when we have access to English training data. However, the same strategy can be applied to both train and test queries if we have the budget of collecting data in target languages.
    \\
    
    - Our findings imply that the back-translation of queries using an intermediate bridge language proves to be a useful strategy in the intent recognition experiments.

\section{Related Work}
We pivot our study of related works into the following buckets - generalised intent and entity detection, entity and intent detection in healthcare, health care in Indian languages and multi-lingual healthcare datasets. 

\noindent \textbf{A) Generalised Intent and Entity Detection Approaches:}
\cite{sun2016online,wang2020active,mu2017streaming,mu2017classification}
focus on detecting novel intents in the form of outlier detection. \cite{mullick2022fine} explore intent classification on legal data. People also work on different detection approaches - few shot \cite{xia2021incremental}, zero shot \cite{xia2018zero}, clustering frameworks \cite{mullick2022framework}.  \cite{yani2022better,sufi2021automated,zhao2021bert} all explore entity detection tasks.  \cite{vanzo2019hierarchical} develop a 
hierarchical multi-task architecture for semantic
parsing sentences for cross-domain spoken dialogue systems. Most of these approaches are very domain and language specific and thus not very useful for the healthcare domain in Indian languages.

\noindent \textbf{B) Entity and Intents in Health Care:} 
 \citet{zhou2021natural} solve different tasks in smart healthcare. 
 \citet{bao2020hhh} 
 build a chat-bot framework using user intents. \citet{bai2022incremental} aim at incremental medical intent detection. 
 \citet{razzaq2017intent,amato2017chatbots} develop an e-Health application using intent-context relationships.  
 \citet{zhang2017bringing} explore medical query intents. 
 Most of the works are done for English and Chinese languages and there is no proper architecture for Indian multilingual scenarios for intent and entity extraction.. 

\noindent \textbf{C) Health Care in Indian Languages:}
Some researchers focus on Indian Languages - Hindi Medical Conversation system, MedBot \cite{bharti2020medbot}, detecting Hindi and English COVID-19 posts \cite{patwa2021overview}, Tamil health information \cite{santosh2019speech},  Bengali health-bot \cite{badlani2021multilingual}, 
 Telugu COVID-19 health information \cite{vishwas2017translation}. But none of the work aims at Indian health query datasets and model analysis.
\cite{mondal2022global} highlights the gaps when using existing state-of-the-art commercial frameworks for NLU tasks in a few Asian and African low-resource languages, especially when the goal is to develop conversational agents for healthcare during COVID. In our work, we strengthen the claims made in their paper for generic healthcare specific datasets in Indian context, and highlight the potential drawbacks of the existing LMs.
 
\noindent \textbf{D) Multilingual Health Care Dataset:} 
\citet{liu2020meddg} develop MedDG (Medical Dialogue dataset of common Gastrointestinal diseases) in Chinese. 
\citet{zeng2020meddialog} proposes MedDialog, a Chinese and English medical dataset, and explores medical dialogue generation tasks. \citet{zhang2021cblue} build a medical intent evaluation dataset in Chinese and \citet{kim2022constructing} has constructed a Korean health intent dataset. 
Our work differs from the existing research in two ways: 1) We focus on developing a gold standard healthcare NLU dataset in Indian languages, 2) cost parameter and availability oriented usage of models for intent detection and entity extraction, and 3) end-to-end evaluation of the state-of-the-art solutions for healthcare in both English and Indic languages which leads to interesting implications and generates important future recommendations for the language community.

\section{Dataset and Pre-Processing}
\label{sec:dataset}
\subsection{Necessity of a new dataset}
India is a country with a diverse language speaking population. 
There is  an increasing population of users consuming Indian language content. 
This linguistic diversity certainly poses significant challenges in healthcare setup, particularly in the situation when healthcare providers and patients speak different languages (also termed as \textit{Language Discordance}) \cite{AlShamsi2020ImplicationsOL}. 
Therefore, individuals with limited English proficiency are left behind and suffer from worse health outcomes than those who speak English with high proficiency. The growing need for the deployment of multilingual conversational agents in hospital and healthcare facilities in India, especially highlighted by the plight of the healthcare workers during the COVID-19 pandemic, warrants a multilingual healthcare query intent dataset in Indian languages \cite{daniel-etal-2019-towards}. Therefore, we resort to create two novel \textbf{I}ndian \textbf{H}ealthcare \textbf{Q}uery \textbf{I}ntent \textbf{D}atasets - (IHQID-WebMD and IHQID-1mg) and one real-world healthcare dataset from hospitals.


\subsection{Source of the dataset}
Due to the unavailability of open-source multilingual NLU datasets in healthcare setup, we sample frequently asked medical queries (FAQs) in English from two popular data sources:
    
\noindent \textbf{WebMD\footnote{\url{https://www.webmd.com/}}:} It is an American website containing a large repository of healthcare data. The queries, taken from the WebMD health forum are asked by ordinary users regarding a wide range of problems. 
\noindent \textbf{1mg\footnote{\url{https://www.1mg.com/}}:} 1mg is an Indian website, which is also a rich source for healthcare data, especially in the Indian context. The English queries are scraped from the FAQ section in drug and disease pages.

Although, both the above datasets are curated from online forums where users post healthcare concerns, in order to evaluate our approach in a practical Indian context, we develop a real world healthcare query dataset in Indian scenario. We collect real world healthcare queries (asked by patients) from the doctors in local hospitals. All queries are anonymous without identity or any details of the patients.
For each language, we fetch 100 queries (some of which overlap) belonging to different categories. 


\subsection{Dataset Sampling} 
The FAQs sampled from these data sources are unlabeled. Hence, for the purpose of supervised classification, it is necessary to categorize each query into a specific intent and list of corresponding entities. 
We broadly categorize queries into four different intent types, namely, \textit{`Disease', `Drug', `Treatment Plan'} and \textit{`Other'}. Each query is assigned one of the four intent labels. Two English-speaking medical graduate doctors annotate the intents 
from the English queries to prepare the datasets. Annotators also mark entities, belong to three different medical entity categories present in the datasets - \textit{`Disease', `Drug'} and \textit{`Treatment'}. The queries with their intent labels are retained where both annotators agree, otherwise discarded. On an average, this filtering lead to an average rejection of around 10\% samples of the dataset for all our setups and languages.
Overall Inter-annotator agreement, Cohen $\kappa$ is 0.89. 

\begin{table*}[t]
\small
\centering
\resizebox{\linewidth}{!}{
\begin{tabular}{lcc|cc|ccccccc}
\toprule
 \multicolumn{3}{c}{\textbf{Intents}} &  \multicolumn{2}{c}{\textbf{Entities}} & \multicolumn{7}{c}{\textbf{Real World Hospital Query Data ($\#$Intent / $\#$Entity)}} \\
\cmidrule(r){1-3}
\cmidrule(r){4-5}
\cmidrule(r){6-12}
 \textbf{Class} & \textbf{$\#$WebMD} & \textbf{$\#$1mg} & \textbf{$\#$WebMD} & \textbf{$\#$1mg} & \textbf{$\#$En} & \textbf{$\#$Hi} & \textbf{$\#$Bn} & \textbf{$\#$Ta} & \textbf{$\#$Te}  & \textbf{$\#$Ma}& \textbf{$\#$Gu}\\
\midrule
 Disease & 283 (207+76) & 111 (87+24) & 629 (464+165) & 240 (185+55)  & 28/37 & 31/35 & 29/37& 27/35 & 31/39 & 28/35& 29/35   \\
 Drug & 234 (181+53)  & 198 (144+54) & 400 (302+98) & 224 (166+58)  &34/44 & 33/43& 31/37& 30/35& 32/38 &34/40 & 32/37 \\
 Treatment & 166 (127+39) & 67 (46+21)  &  218 (165+53)  & 64 (44+20)  & 21/24& 20/26  &  21/25& 23/29 & 19/24& 17/23 & 20/26\\
  Other & 278 (205+73)  & 41 (28+13)  & - & - &17/- & 16/- & 19/-& 20/-& 18/-  & 21/- & 19/-\\
 \midrule
 \textbf{Total} & 961 (720+241) & 417 (305+112)  & 1247 (931+316) & 528 (395+133) & 100/105 &100/104 &100/99 &100/99 &100/101 &100/98&100/98 \\ 
\bottomrule
\end{tabular}
}
\caption{Distributions of different types of intent and entity labels in WebMD, 1mg datasets (IHQID) and Real World Hospital Query Data. (- + -) represents (train + test) division. $\#$ denotes the count. }
\label{tab:dataset}
\end{table*}

\subsection{Parallel Data Generation}
In order to generate parallel corpora of these frequently asked questions in English, we choose six Indian languages apart from English.\\
\textbf{Language Selection:}
The language set includes English: USA version (EN-US) termed as (`En'), Hindi (`Hi'), Bengali (`Bn'), Tamil (`Ta'), Telugu (`Te'), Gujarati (`Gu') and Marathi (`Mr'). The choice of languages was driven by (a) the number of native speakers of those languages in India, b) number of annotators available for creating the dataset, (c) combined with typological diversity amongst the languages - we choose languages from various language families. For instance, Bengali, Hindi, Gujarati, Marathi belong to the Indo-Aryan family whereas Tamil and Telugu belong to the Dravidian group.\\
\noindent
\textbf{Annotation and Quality Control:}
Since the gold standard annotated queries are not available online in Indian languages, the English queries of 1mg and WebMD have to be manually translated. \textbf{After discussions with the doctors and different patients, we create the annotation guidelines.}
Annotators are told to formulate the queries on their own regional languages with the help of Bing Translator API\footnote{ \url{https://www.microsoft.com/en-us/translator/business/translator-api/}}. Annotators are also asked to annotate the entities and their types (in their respective native languages) for each query being corrected with the idea of what common people of corresponding native language generally ask healthcare queries to doctors.

Three annotators are selected per language after several discussions and conditions of fulfilling many criteria like annotators should have native proficiency in their language of annotation, domain knowledge expertise along with a good working proficiency in English. Initial labeling is done by two annotators and any annotation discrepancy is checked and resolved by the third annotator after discussing with others. 
While formulating the query on their own manually, the annotators are also asked to annotate the entities and their types (in their respective native languages) for each query being corrected. The above quality control measures ensure that the translated data is of high quality, resembling real world data in the target language. 
In the case of a word such as a proper noun like \textit{`Paracetamol'} (drug), which does not have a translation in the target vocabulary, the word is asked to be simply transliterated in the target language.

 In order to prepare the real world hospital query dataset in Indian healthcare contexts, we collect healthcare queries from the doctors of local hospitals. It also consists of six different Indic languages along with English. There are a hundred queries for each of the language. These queries also have similar intent classes and entity categories, which are labelled by the doctors. 
 During collection of queries, we fix the minimum number of samples for each intent classes across all languages.
 
In order to maintain the quality of the Indian language annotations, the annotators are directed to use the native language words and grammar, keeping the original interpretation of the query. 
All query logs, annotations and changes are recorded in order to conduct future verification and analysis. On completion of the translation process, the annotators are asked to exchange their work and check the quality of translation for fluency and semantic stability. 
Inaccuracies are noted, and the respective queries are rectified in the dataset.

At the end, we finally have three multilingual intent and entity recognition labelled datasets - \textbf{IHQID-WebMD}, \textbf{IHQID-1mg} and a real world hospital query test dataset 
in seven different languages, the dataset distributions of which are provided in Table \ref{tab:dataset}. The first two datasets (IHQID-WebMD and IHQID-1mg) help to build the models and real world hospital dataset is used to evaluate our approaches in real world contexts. Table \ref{tab:dataset} also shows the statistical details across different intent classes (`Disease', `Drug', `Treatment' and `Other') and corresponding entities (of `Disease', `Drug' and `Treatment' categories) along with the total counts and train-test divisions. It also shows the distribution of hospital collected practical healthcare queries across different languages (Right part of the table).

\section{Strategies of Evaluation }
In this section,  we illustrate the strategies of evaluating the state-of-the-art LMs on our dataset.
Our evaluation of these models for Healthcare is scoped down to two fundamental NLU tasks: \\
a) Intent Recognition (Section 5.1) \\
b) Entity Extraction (Section 5.2)

\noindent
\noindent \textbf{Evaluation Setup Description: }
Our evaluation of the models has been conducted while keeping in mind about the availability of human-translated monolingual and multilingual training data in two possible real-life scenarios: 1) \textbf{Scenario A:} In this setup, we assume to have access to only English training data (less costly) and in 2) \textbf{Scenario B:} we assume to have access to manually written training queries in all the target languages (very expensive).
During inference/testing, we expect all the queries are in the corresponding target languages.
\noindent
\underline{\textbf{Scenario A:}} \\
\noindent \textbf{Setup 1) Backtranslated Test (S1): [Translate-Test]} Here we develop our system by training the models on the English queries, and evaluate the intent detection and entity extraction systems in different languages by automatically backtranslating the test queries into English (e.g. similar to \cite{gupta2021truthbot}). 
\noindent
\textbf{Setup 2) Zero-Shot Cross-Lingual Test (S2):} 
Cross lingual transfer learning is a useful methodology used for tasks involving scarce data \cite{zhou-etal-2016-cross, karamanolakis-etal-2020-cross}.
In this setup,
the models make use of zero-shot based cross-lingual capabilities from training on the English data (scraped from WebMD and 1mg) and use it for inference on test queries in Indic languages. 
\noindent
\textbf{Setup 3) Bridge Language Backtranslation (S3):} 
Here a relatively low-resource language is first translated to an intermediate language and then finally to English. The motivation behind this setup lies in the fact that even though these Indic languages belong to different scripts, there are linguistic and morphological similarities among them which may improve the translation to English if they are used as intermediate languages. In this paper, we have considered `Hindi' as the bridge language. This notion of such ``bridge" languages has been explored previously in the context of Machine Translation \cite{mt_pivots} and zero/few-shot transfer in MMLMs \cite{lauscher-etal-2020-zero}. \\\
\noindent
\underline{\textbf{Scenario B:}}\\
\noindent \textbf{Setup 4) Train and Test on Indic Data (S4):} 
In this setup, we use the training dataset in indic languages to train our NLU models in different target languages. Here, we use the 
IHQID-WebMD and IHQID-1mg Indic data (non-English)
to evaluate the NLU detection performances of the developed models. Jennifer Bot \cite{li-etal-2020-jennifer} use a similar setup to extend their English bot to Spanish. 
\noindent \textbf{Setup 5) Full Backtranslation (S5):} In this setup, both train and test data are backtranslated to English.
This is useful for the countries with poor technical setups for low-resource languages, since an automated approach can translate low-resource medical queries to resource-rich language and test.

In all back translation experiments, we use Bing Translation Api \footnote{\url{https://www.microsoft.com/en-us/ translator/business/translator-api/}}.

\aboverulesep=0ex
\belowrulesep=0ex
\renewcommand{\arraystretch}{1.1}

\section{Experiments and Results}
\textbf{Experimental Setup:}
Our experiments are conducted on two Tesla P100 GPUs with 16 GB RAM, 6 Gbps clock cycle and GDDR5 memory. All methods of entity extraction and intent detection took less than 30 GPU minutes for training. 
We perform a hyperparameter search and report the results of the settings which achieve the best results, and then fixed the same for all the models. 
The batch size is kept at 16, number of epochs is 10, optimization algorithm used is AdamW and the learning rate is 1e-5 with cross-entropy as the loss function. 

\subsection{Intent Detection}
\textbf{Task Description:} It can be defined as a multi-class classification task of correctly assigning a medical query with an intent label from a fixed set of intents (\textit{drug, disease, treatment} and \textit{other}).

\noindent \textbf{Classification Models:}
Since in Setups 1, 3 and 5, we take both the training and test set in English, we use state-of-the-art LMs pre-trained on English corpora 
(as shown in (i)) for our classification experiments. 
Whereas in Setup 2 and 4, we make use of multilingual LMs  (as shown in (ii)) which have been widely used for various benchmark tasks in Indian languages.  Following are the baselines:

\textbf{(i) Pre-trained English Models:}
For setups 1, 3 and 5, we fine-tune the last layer of RoBERTa \cite{liu2019roberta} and Bio\_ClinicalBERT \cite{alsentzer2019publicly} models on the English queries for intent detection by adding a classification layer that takes [CLS] token as input. The latter is a state-of-the-art domain-specific transformer based language model pre-trained on MIMIC III notes\footnote{\url{https://huggingface.co/emilyalsentzer/Bio_ClinicalBERT}}, which is a collection of electronic health records and discharge notes. 

\textbf{(ii) Pre-trained Multilingual Models:}
Two pre-trained mulilingual LMs are used, mBERT \textit{(bert-base-multilingual-uncased)} \cite{pires-etal-2019-multilingual} and XLM-Roberta \textit{(xlm-roberta-base)} \cite{conneau-etal-2020-unsupervised}, both support all Indic languages in the datasets along with English. 
In Setup 2, we perform zero-shot classification using these models.
The zero shot setting involves fine-tuning the model using English data, and testing on Indic languages.
Whereas in Setup 4, we first train these models using the entire train sets in the target languages, separately for WebMD and 1mg, and check the performance on the test sets. 


\subsection{Entity Recognition}
\noindent \textbf{Task Description:}
This task is analogous to performing a Named Entity Recognition (NER) for three categories, namely, \textit{drugs, diseases and treatments} on the query texts. We follow the standard BIO-tagging system while annotating the entities word-by-word. The train and test files for each configuration and language respectively are constructed from our WebMD and 1mg datasets.


\noindent \textbf{Extraction Frameworks:}
For entity recognition, we follow the same strategies of evaluating the predictive performance of the LMs as described in Section 4.
The same models (as described in section 5.1) are also used for entity recognition experiments. 

\input{tables/intent_shuffled}
\subsection{Evaluation}
For all our experiments on intent detection and entity recognition, we calculate the Precision, Recall and report the F1-score. 

\subsection{Results and Analysis}
\underline{\textbf{Intent Detection:}} Table \ref{tab:intents-A} shows the results of intent detection of five experimental strategies on the IHQID-WebMD and IHQID-1mg datasets in terms of Macro F1-score (in percentage). 

\noindent
\textbf{Finding 1:} We observe that in general, Backtranslated Test (Setup 1) performs better than Zero-Shot Cross-Lingual Test (Setup 2).
Moreover, it is interesting to notice that even though the performance of these models for most of the target languages in Setup 1 are comparable with that of English in WebMD (an average of 3\% drop for all the languages compared to English), there is a significant drop (average of 6\%) in the F1 scores for the Setup 1 results in 1mg Dataset. This holds true for both RoBERTa and BcBERT experiments.
This 
denotes that the state-of-the-art English models, which are performing decently after backtranslation of the medical queries in English, pre-trained on both generic and medical domain, are lagging behind when the vocabularies of the medical entities are in the Indian context.
This definitely calls for an immediate attention to developing LMs pre-trained on India-specific medical datasets.

\input{tables/entity_shuffled}
\noindent
\textbf{Finding 2:} Another interesting observation was that the use of Bridge Language Backtranslation (Setup 3) in Table~\ref{tab:intents-A}, helps to boost performance of most of the languages in the case of 1mg dataset in comparison to Setup 1. The observation does not hold true for intent recognition in WebMD dataset. 
This might be attributed to the fact that using a bridge Indian language as an intermediate helps preserve the domain-specific sense of the queries instead of directly converting the queries from the target language to English. This seems like a reasonable alternative to develop useful intent recognition models for healthcare in Indian languages. 

\noindent
\textbf{Finding 3:} In comparison with zero-shot cross-lingual transfer (Setup 2), both mBERT and XLM-R models are outperformed by few-shot experiments (Setup 4) for intent detection. This observation holds true for both WebMD and 1mg datasets. However, Setup 4 is much more cost-intensive than the Setup 2.

\noindent \textbf{Finding 4:} We report the average (Avg) F1-score across all languages. The best performing model is RoBERTa (Setup 1 for English and Setup 5 for non-English) for both WebMD (74.94\%) and 1mg (70.33\%). RoBERTa is used for further evaluations.\\

\begin{figure*}[htb]
\centering
        \begin{subfigure}[b]{0.24\textwidth}
                \centering            \includegraphics[width=.95\linewidth]{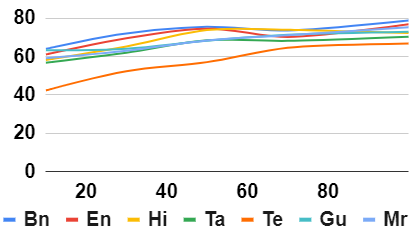}
                \caption{Intent - WebMD}
                \label{fig:Intent-WebMD}
        \end{subfigure}%
        \begin{subfigure}[b]{0.24\textwidth}
                \centering \includegraphics[width=.95\linewidth]{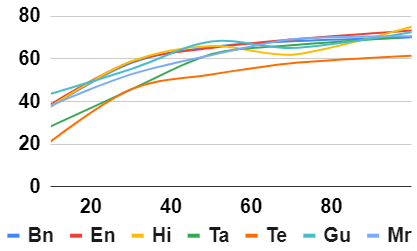}
                \caption{Intent - 1mg}
                \label{fig:Intent-1mg}
        \end{subfigure}%
        \begin{subfigure}[b]{0.24\textwidth}
                \centering            \includegraphics[width=.95\linewidth]{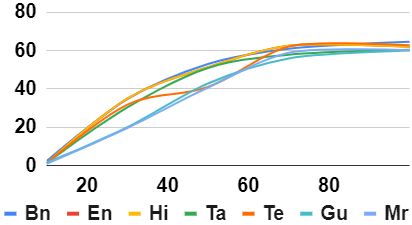}
                \caption{Entity - WebMD}
                \label{fig:Entity-WebMD}
        \end{subfigure}%
        \begin{subfigure}[b]{0.24\textwidth}
                \centering \includegraphics[width=.95\linewidth]{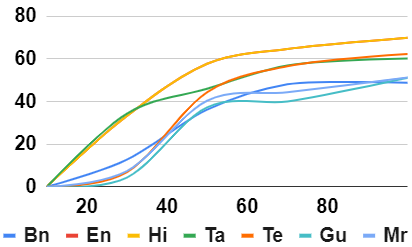}
                \caption{Entity - 1mg}
                \label{fig:Entity-1mg}
        \end{subfigure}%
\caption{Intent Detection and Entity Extraction F1-score (Y-axis) for Different Percentage of Training Data (X-axis) for WebMD and 1mg} 
        \label{fig:intent-entity}
\end{figure*}


\noindent\underline{\textbf{Entity Extraction:}}
Table~\ref{tab:entity-A} displays the results of entity recognition task under five different strategies on IHQID-WebMD and IHQID-1mg datasets. \\
\noindent \textbf{Finding 1:} In the Backtranslation test performed in Setup 1, we observe that for WebMD dataset, the difference in the performance of the models (Performance on English is 0.33\% more average F1 Score for RoBERTa and 3.58\% more than average F1 for bcBERT) is far less significant than the drop observed for 1 mg (Performance on English is 9.66\% more average F1 Score for RoBERTa and 10.49\% more than average for bcBERT). This implies that loss of information is quite high for the entities in Indian context during backtranslation.\\
\noindent
\textbf{Finding 2:} Unlike our findings on Setup 3 in intent recognition, we observe that backtranslation using a bridge language seems to induce more loss of information on the entities compared to Setup 1. This observations holds true for both the models across two datasets. \\
\noindent
\textbf{Finding 3:} 
Similar to intent recognition, we observe that completely backtranslating both training and test data to English performs the best among S1, S3 and S5. This holds true for both the datasets and both the models. However, this operation is indeed expensive in terms of data curation cost, since it requires original data in the target languages for both training and testing.\\
\noindent
\textbf{Finding 4:}
The abysmal performances of the multilingual models as shown in Table~\ref{tab:entity-A}, for both S2 and S4 indicate that these approaches are not so useful in our case.  

\noindent \textbf{Finding 5:} We report the average (Avg) F1-score across all languages. BioClinicalBERT performs the best (Setup 1 for English and Setup 5 for non-English (Avg)) for both WebMD (63.14\%) and 1mg (68.69\%). It is used for further evaluations.\\

\subsection{Ablation Study}
\noindent \textbf{Experiments with Varying Training Size:} We experiment with varying training sizes on both intent detection and entity extraction tasks using the best performing models, by taking 10\%, 30\%, 50\%, 70\% and 100\%) of the training set. We then show the F1-scores (Y-axis) for all the languages with different training sizes (X-axis) in Fig. \ref{fig:intent-entity}. Fig. \ref{fig:Intent-WebMD} and \ref{fig:Intent-1mg} show that the performance of the intent detection models do not vary too much with increasing training sample data. However, Fig. \ref{fig:Entity-WebMD} and \ref{fig:Entity-1mg} clearly show that entity extraction F1-scores increase significantly with the increase of training data. Thus, we can conclude that the intent detection model does not require a large amount of data to generalise, as opposed to the requirements of the entity extraction model.

\begin{table*}[t]
\centering
\resizebox{0.95\textwidth}{!}{
\begin{tabular}{| l | cccc | cccc | ccc | ccc |}
\toprule
\multirow{3}{*}{Lang} &\multicolumn{8}{c|}{Intent} &\multicolumn{6}{c|}{Entity} \\\cmidrule{2-15}
&\multicolumn{4}{c|}{WebMD} &\multicolumn{4}{c|}{1mg} &\multicolumn{3}{c|}{WebMD} &\multicolumn{3}{c|}{1mg} \\\cmidrule{2-15}
&Disease &Drug &Treatment &Other &Disease &Drug &Treatment &Other &Disease &Drug &Treatment &Disease &Drug &Treatment \\\midrule
En &75.86 &81.42 &74.16 &74.07 &80.00 &94.64 &85.00 &35.29 &63.16 &72.13 &61.39 &66.67 &88.00 &47.06 \\
Hi &73.10 &80.00 &66.67 &69.50 &72.97 &78.57 &67.47 &69.06 &64.37 &70.41 &58.72 &67.27 &85.04 &55.56 \\
Bn &80.79 &80.39 &71.91 &75.71 &80.77 &94.74 &87.18 &52.63 &65.19 &69.35 &53.23 &73.50 &68.48 &47.72 \\
Ta &77.63 &73.27 &60.00 &73.38 &80.77 &94.74 &80.95 &25.00 &60.05 &69.07 &49.23 &76.11 &72.87 &50.00 \\
Te &72.85 &78.10 &70.45 &72.46 &80.77 &94.55 &77.27 &33.33 &62.09 &65.00 &60.71 &75.63 &65.37 &66.67 \\
Gu &75.64 &78.50 &69.77 &72.18 &83.02 &93.81 &80.00 &33.33 &53.41 &66.32 &58.82 &72.41 &69.44 &52.86 \\
Mr &76.82 &78.85 &75.29 &71.83 &79.25 &94.64 &83.72 &25.00 &59.48 &59.26 &56.45 &60.78 &59.74 &49.58 \\
\bottomrule
\end{tabular}}
\caption{Macro-F1 scores for intent identification and entity extraction on the WebMD (WMD) and 1mg datasets.  For each language, we portray the results of the best model obtained for the corresponding dataset.}
\label{tab:intents-entity-category}
\end{table*}





\noindent \textbf{Category wise intent detection and entity extraction for the best model:} We evaluate the F1-scores for different intent classes for the RoBERTa Model (Setup 1 for English and Setup 5 for non-English) trained on WebMD and 1mg (See Section 4 for setup descriptions). Similarly, with the help of BioClinicalBERT (Setup 1 for English and Setup 5 for Non-English), we find the individual entity class wise F1-scores. The results in Table \ref{tab:intents-entity-category} 
 show that the model is able to detect `disease', `drug' and `treatment' intent classes with high F1-score but 
 the performance on the `Other' class is poor, thus bringing the macro averaged F1 score down considerably.
 This may be due to the fact that the system fails to detect open ended query types, present in the `Other' class. This is supported by the intent class wise entity distribution, which shows an overwhelming dominance of `drug', `disease' and `treatment' entities in their corresponding intent categories (`drug', `disease' and `treatment plan' intents, respectively), whereas the `other' intent class, of which there are very few instances comparatively anyway, has no such dominant entity class associated with it.
 In the entity extraction task, the best performing model is able to extract all three entity categories with a similar F1-score performance.

\begin{figure}[!htp]
    \centering
    \begin{adjustbox}{width=0.85\linewidth}
\includegraphics[width=\linewidth]{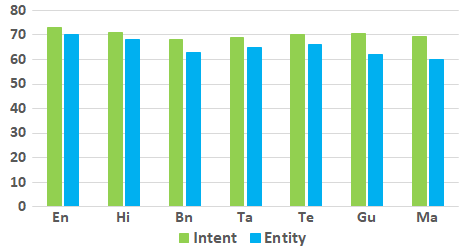}
    \end{adjustbox}
    \caption{Macro Average F1 Score for Intent detection and Entity Extraction across all different languages in Real World Hospital Data}
    \label{fig:hospital-data}
\end{figure}

\noindent \textbf{Real World Hospital Data Evaluation:}
We use the real world healthcare query dataset (100 queries per language) to test the usability of our models in practical Indian hospital scenarios. We run the best performing models trained on WebMD and 1mg data for intent detection (RoBERTa in Setup 1 for English and Setup 5 for Non-English) and entity extraction (BioCLinicalBERT in Setup 1 for English and Setup 5 for non-English) and report the average of two models (trained on IHQID-WebMD and IHQID-1mg) for each language. Fig. \ref{fig:hospital-data} shows the
average F1-score for each language,
which is consistent with the earlier results shown in Table \ref{tab:intents-A} and \ref{tab:entity-A}. This shows that the best performing proposed setup performs satisfactorily on real world data in Indic languages.


\subsection{Demonstration} 
To be able to make the proposed methods accessible and usable by the community, we create an online interface, which could be found in our GitHub repository\footnote{\href{https://github.com/indichealth/indic-health-demo}{https://github.com/indichealth/indic-health-demo}}. With the help of this website, one can post health query in the allowed language and obtain the predictions using our best
models. 
\section{Discussion and Error Analysis} 

\begin{figure} 
\vspace{-2mm}
    \centering
    \includegraphics[width=\linewidth]{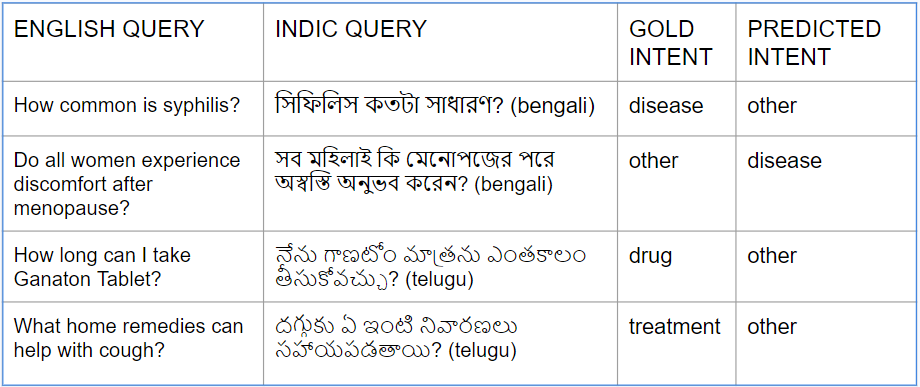}
    \vspace{-4mm}
    \caption{Error in Prediction}
    \vspace{-2mm}
    \label{fig:error-pred}
\end{figure}

\begin{figure} 
    \centering
    \includegraphics[width=\linewidth]{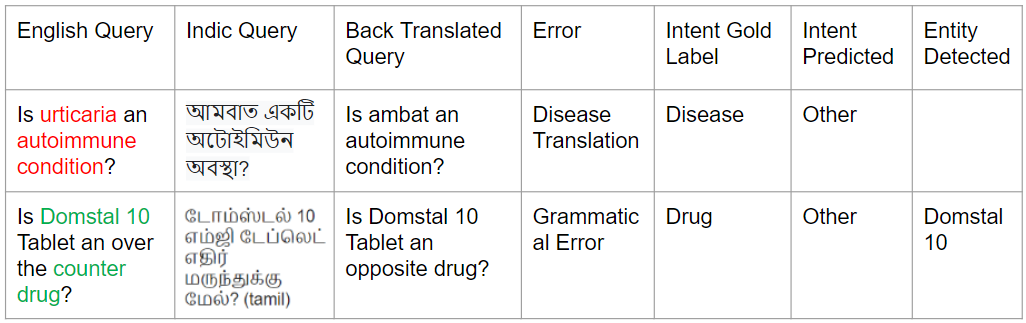}
    \caption{Error in Back-Translation}
    \vspace{-4mm}
    \label{fig:error-translation}
\end{figure}

We categorize the issues in mis-classification and identify two broad themes of the reasons.
The primary reason is model prediction error. Figure \ref{fig:error-pred} shows the model prediction errors for various intents in different languages. For an example, `How common is syphilis' is of `disease' intent category but model wrongly predicts it as `other' category. 
Another reason is the misclassification due to incorrect translation of the medical entities such as the disease \textit{`uticartia'} has been transformed into \textit{`ambat'} during backtranslation as shown in Figure \ref{fig:error-translation} which is not detected as an entity. 
So, the backtranslation error leads to intent mis-classification and entity extraction error. 
We speculate such random absurd behaviour due to the context of the query and languages are semantically different. Secondly, there are also certain issues in fluency and grammatical meaning after backtranslation. For instance, \textit{`over the counter drug'} gets changed to \textit{`over the opposite drug'}. Entity recognition errors are also occurring along with the intent mis-classification.

\section{Conclusion}
\label{sec:reco}
We focus on developing novel Indian HealthCare Query Datasets and propose frameworks to detect intents and extract entities from queries in different Indian languages. 
Through extensive experiments on our proposed datasets,  we recommend the community to use backtranslation of test queries to English in two real-life scenarios as a reasonable choice when we have access to English training data. However, the same strategy can be applied to both train and test queries if we have the budget of collecting data in target languages.
Backtranslation of queries using an intermediate bridge language also proves to be a useful strategy in some cases.

\section*{Acknowledgements}
 The project was supported in part by the grant given by I-Hub Foundation for Cobotics, IIT Delhi for the project, "Voice based Natural Interaction for Goal Oriented Tasks in Healthcare".

\section*{Limitations}
Our dataset needs to be scaled up in terms of size and intent labels which we aim to do as a part of future work. Another constraint is that we do not consider cases where queries are multi-labelled (e.g. - drug and disease both). We shall explore in future.

\section*{Ethical Concerns}
We propose to release the dataset which neither reveals any personal sensitive information of the patients nor any toxic statement. Besides, we have paid enough token money (exact remuneration will be revealed once accepted to the conference) to the domain-expert annotators who have helped us in manually tagging the medical queries. 

\bibliography{anthology,custom}
\bibliographystyle{acl_natbib}

\appendix



\end{document}

%% file: tables/intent_shuffled.tex
\begin{table*}[t]
\centering
\resizebox{\textwidth}{!}{
\begin{tabular}{| l | cccc | cccc | cccc | cccc | cccc |}
\toprule
& \multicolumn{4}{c|}{Backtranslated Test (S1)}  &\multicolumn{4}{c|}{Zero-Shot Cross-Lingual Test (S2)} &\multicolumn{4}{c|}{Bridge Language Backtranslation (S3)} &\multicolumn{4}{c|}{Train and Test on Indic Data (S4)} &\multicolumn{4}{c|}{Full Backtranslation (S5)}\\
\cmidrule{2-21}
Lang & \multicolumn{2}{c}{RoBERTa} &\multicolumn{2}{c|}{bcBERT}  &\multicolumn{2}{c}{mBERT} &\multicolumn{2}{c|}{XLM-RoBERTa} &\multicolumn{2}{c}{RoBERTa} &\multicolumn{2}{c|}{bcBERT} &\multicolumn{2}{c}{mBERT} &\multicolumn{2}{c|}{XLM-RoBERTa} &\multicolumn{2}{c}{RoBERTa} &\multicolumn{2}{c|}{bcBERT} \\
\cmidrule(r){2-3}
\cmidrule(r){4-5}
\cmidrule(r){6-7}
\cmidrule(r){8-9}
\cmidrule(r){10-11}
\cmidrule(r){12-13}
\cmidrule(r){14-15}
\cmidrule(r){16-17}
\cmidrule(r){18-19}
\cmidrule(r){20-21}
-uage&WMD &1mg &WMD &1mg &WMD &1mg &WMD &1mg &WMD &1mg &WMD &1mg &WMD &1mg &WMD &1mg &WMD &1mg &WMD &1mg \\\midrule
En &\underline{76.34} &\underline{73.33}  &75.38 &68.72 &- &- &- &- &- &- &- &- &- &- &- &- &- &- &- &- \\
Hi &73.90 &67.48 &66.50  &66.50   &42.46 &46.45 &58.68 &43.30  &- &- &- &- &56.18 &51.41 &41.14 &40.09 &\underline{75.42} &\underline{72.32} &75.21 &63.81 \\
Bn &75.18  &66.42  &75.02  &63.66  &35.85 &35.62 &55.85 &43.69 &70.76 &\underline{71.94} &71.91 &64.87 &50.26 &46.65 &41.07 &39.73 &\underline{78.83} &70.13 &75.41 &57.52 \\
Ta &73.63 &64.29 &73.99 &62.88 &38.50 &39.34 &57.47 &42.14 &69.51 &66.42 &73.06 &64.43 &51.17 &50.49 &40.04 &32.63  &\underline{74.36} &\underline{69.44} &73.79 &62.82 \\
Te &73.25 &63.48 &73.79 &66.17 &36.40 &30.75 &55.38 &38.53 &69.89 &66.63 &72.19 &63.67 &51.28 &51.69 &45.26 &41.07 &71.80 &\underline{66.90}  &\underline{75.30} &65.63 \\
Gu &71.76 &66.85 &73.05 &68.80 &35.61 &32.67 &51.50 &34.58 &71.61 &72.07 &\underline{73.75} &68.02 &50.07 &51.17 &42.23 &46.23 &72.93 &\underline{72.54} &71.76 &70.52  \\ 
Mr &72.70 &70.64 &73.47 &73.26 &43.57 &38.22 &60.58 &44.16 &71.50 &72.47 &73.13 &\underline{74.28} &54.44 &55.24 &43.18 &46.11 &\underline{76.32} &70.65 &75.42 &63.83  \\\hline
Avg & 73.82 & 67.50 & 73.03 & 67.14 & 38.73  & 37.18 & 56.58 & 41.07 & 70.65 & 69.91  & 67.25 & 67.05 & 52.23  & 51.10 &  42.15 & 40.98 & \underline{74.94} & \underline{70.33} & 74.48 & 64.03\\
\bottomrule
\end{tabular}}
\caption{Macro-F1 scores for intent classification on the WebMD (WMD) and 1mg datasets for five Setups (three different setups for Train on English (Scenario A) and two setups of Train on Indic Data (Scenario B)). 
bcBert indicates BioClinicalBERT, mBERT indicates Multilingual BERT. 
\underline{Underline} denotes the best across five settings.}
\label{tab:intents-A}
\end{table*}

%% file: tables/entity_shuffled.tex
\begin{table*}[bp]
\centering
\resizebox{\textwidth}{!}{
\begin{tabular}{| l | cccc | cccc | cccc | cccc | cccc |}
\toprule
& \multicolumn{4}{c|}{Backtranslated Test (S1)}  &\multicolumn{4}{c|}{Zero-Shot Cross-Lingual Test (S2)} &\multicolumn{4}{c|}{Bridge Language Backtranslation (S3)} &\multicolumn{4}{c|}{Train and Test on Indic Data (S4)} &\multicolumn{4}{c|}{Full Backtranslation (S5)}\\
\cmidrule{2-21}
Lang & \multicolumn{2}{c}{RoBERTa} &\multicolumn{2}{c|}{bcBERT}  &\multicolumn{2}{c}{mBERT} &\multicolumn{2}{c|}{XLM-RoBERTa} &\multicolumn{2}{c}{RoBERTa} &\multicolumn{2}{c|}{bcBERT} &\multicolumn{2}{c}{mBERT} &\multicolumn{2}{c|}{XLM-RoBERTa} &\multicolumn{2}{c}{RoBERTa} &\multicolumn{2}{c|}{bcBERT}  \\
\cmidrule(r){2-3}
\cmidrule(r){4-5}
\cmidrule(r){6-7}
\cmidrule(r){8-9}
\cmidrule(r){10-11}
\cmidrule(r){12-13}
\cmidrule(r){14-15}
\cmidrule(r){16-17}
\cmidrule(r){18-19}
\cmidrule(r){20-21}
-uage &WMD &1mg &WMD &1mg &WMD &1mg &WMD &1mg &WMD &1mg &WMD &1mg &WMD &1mg &WMD &1mg &WMD &1mg &WMD &1mg \\\midrule
En &61.95 &69.93 &\underline{65.50} &\underline{73.68} &- &- &- &- &- &- &- &- &- &- &- &- &- &- &- &- \\
Hi &61.75 &69.58 &65.20 &73.82 &34.75 &34.01 &36.53 &46.95  &- &- &- &- &17.55 &36.85 &52.43 &71.68 &60.60 &69.32 &\underline{65.90} &\underline{76.55} \\
Bn &64.21 &56.79 &64.25 &62.56 &35.12 &35.02 &34.31 &41.42 &61.05 &47.45 &59.70 &50.08 &27.28 &42.37 &63.73 &62.69 &64.47 &54.81 &\underline{64.75} &\underline{65.80} \\
Ta &60.44 &60.58 &60.22 &60.72 &30.62 &30.10 &34.31 &36.29 &55.35 &56.75 &56.28 &63.48 &22.07 &29.87 &59.91 &67.59 &61.07 &69.63 &\underline{64.91} &\underline{71.40} \\
Te &62.76 &62.56 &62.37 &63.44 &30.00 &31.50 &32.95 &41.71 &62.26 &55.75 &63.89 &62.27 &27.95 &27.82 &59.81 &68.93 &65.27 &67.06 &\underline{66.19} &\underline{69.17} \\
Gu &60.02 &51.13 &58.20 &52.62 &23.56 &27.24 &23.90 &42.19 &56.36 &47.12 &57.60 &51.47 &21.93 &25.82 &49.19 &\underline{73.77} &\underline{60.78} &59.62 &58.26 &70.78 \\
Mr &\underline{60.18} &51.31 &57.68 &55.45 &26.32 &22.54 &29.51 &50.84 &54.63 &59.61 &55.02 &\underline{60.96} &20.48 &23.52 &52.61 &57.56 &59.10 &57.38 &58.83 &58.45 \\\hline
Avg & 61.62& 60.27& 61.92& 63.19&30.56  &30.07& 31.92& 43.23 & 57.92& 53.34& 58.49& 57.65&22.88 & 26.61&  58.28&67.04 & 61.88& 62.98& \underline{63.14}& \underline{68.69}\\
\bottomrule
\end{tabular}}
\caption{Macro-F1 scores for entity extraction on the WebMD (WDM) and 1mg datasets for five Setups (three different setups for Train on English (Scenario A) and two setups of Train on Indic Data (Scenario B)). 
bcBert indicates BioClinicalBERT, mBERT indicates Multilingual BERT. 
\underline{Underline} denotes the best across five settings.}
\label{tab:entity-A}
\end{table*}